\documentclass[10pt,twocolumn]{article}
\usepackage[margin=1in]{geometry}
\usepackage{amsmath}
\usepackage{abstract}
\usepackage{amsfonts,color}
\usepackage{amssymb}
\usepackage{graphicx}
\usepackage{epstopdf}

\begin{document}
\title{\bf{A parametric model to study the mass radius relationship of stars}}
\date{}
\author
{Safiqul Islam$^\ast$, Satadal Datta$^\dagger$ and Tapas K Das $^\ddagger$\\
Harish-Chandra research Institute, Chhatnag Road, Jhunsi, Allahabad-211019, India \\
Homi Bhabha National Institute, Training School Complex, \\Anushaktinagar, Mumbai - 400094, India\\
$^\ast$safiqulislam@hri.res.in\\ $^\dagger$satadaldatta@hri.res.in\\$^\ddagger$tapas@hri.res.in
}
\twocolumn[
\maketitle
\begin{onecolabstract}
In static and spherically symmetric spacetime, we solve the Einstein Maxwell equations. The effective gravitational potential and the electric field for charged anisotropic fluid are defined in terms of two free parameters. For such configuration, the mass of the star as a function of stellar radius is found in terms of two aforementioned parameters, subjected to certain stability criteria. For various values of these two parameters one finds that such mass radius relationship can model stellar objects located at various regions of Hertzsprung-Russel diagram. 
\end{onecolabstract}
]
\vspace*{0.33cm}
\section{Introduction:}
~~~~For relativistic charged fluid with the signature of pressure anisotropy, where the anisotropy is defined by the finite non zero difference between the radial and the tangential fluid pressure, the Einstein Maxwell field equations are solved for static spherically symmetric spacetime. Certain functional form of the electric field as well as the effective gravitational potential have been introduced in our model, where such field and potential are characterized by two free parameters a and b, with certain relationships defined between these two parameters, where such relationships are obtained using a particular form of stability criteria. The charge and the mass energy density have been expressed (as a consequence of the interior solution) as a function of the radial distance. From there, we obtain the mass-radius relationship for the interior solution. Once such mass-radius relationship is integrated for a particular limit defined by the radius of the star, one can obtain what will be the mass of the charged\\\\\\\\\\\\\\ fluid considered in our model, embedded within a sphere of radius R. Hence our model here  provides the mass M(R) of star of radius R. M(R) in our calculations, however, is characterized by (a,b), and there remains a specific relationship between a and b, which are obtained by using some predefined stability criterion. Various values of a and b provides various [M(R)-R] measurements. For different values of a and b, one can find M(R) for different values of R, and hence using our model, we can study the mass radius relation for different categories of stellar objects located at various regions of the Hertzsprung-Russel diagram.
\section{Einstein-Maxwell equations:}

We consider the interior spacetime of a $(3+1)$-D star in Schwarzschild coordinates $(t,r,\theta,\phi)$ as

\begin{equation}
ds^{2}= -e^{2\nu(r)}dt^{2}+e^{2\lambda(r)}dr^{2}+r^{2}(d\theta^{2}+sin^{2}\theta
d\phi^{2})
\end{equation}

Here $\nu$ and $\lambda$ are
the metric potentials which have functional dependence on the radial coordinate r and
$\nu(r)$ is to be determined.

The Hilbert action coupled to electromagnetism is given by
\begin{equation}
I = \int d x^3 \sqrt{-g } \left( \frac{R}{16 \pi}
-\frac{1}{4} F_a^c F_{bc} + L_{m} \right),
\end{equation}

where $L_{m}$ is the Lagrangian for matter. The variation with respect to the
metric gives the following self consistent Einstein-Maxwell equations
for a charged anisotropic fluid distribution,

\begin{eqnarray}
G_{ab}&=&R_{ab} - \frac{1}{2} R g_{ab} = - 8 \pi T_{ab} \nonumber\\
      &=&- 8 \pi (T_{ab}^{(m)} +T_{ab}^{EM}),
\end{eqnarray}

The explicit forms of the energy momentum tensor (EMT) components
for the matter source (we assumed that the matter distribution at
the interior of the star is anisotropic) and
electromagnetic fields are given by,

\begin{equation}
T_{ab}^{(m)} = (\rho +p_{t}) u_au_b - p_{t}g_{ab} + (p_{r}-p_{t}) v_{a}v_{b},
\end{equation}

\begin{equation}
T_{ab}^{EM} = -\frac{1}{4 \pi } \left( F_a^c F_{bc} -\frac{1}{4}
g_{ab} F_{cd}F^{cd}\right),
\end{equation}

where $\rho$, $p_{r}$, $p_{t}$, $u_a$, $v_{a}$ and $F_{ab}$ are, respectively,
matter-energy density, radial fluid pressure, transverse fluid pressure, four velocity, radial four vector of
the fluid element and electromagnetic field tensor. The case $p_{t}=p_{r}$, corresponds to the isotropic fluid when the anisotropic force vanishes. We also consider $G=c=1$ in our observations.

In our consideration, the four velocity and radial four vector satisfy,
$u^{a} = e^{-\nu}{\delta^a_{0}}$, $u^{a} u_{a}=1$, $v^{a} = e^{-\lambda}{\delta^a_{1}}$, $v^{a} v_{a}=-1$.

Also, the electromagnetic field is related to current four vector as,

\begin{equation}
J^c = \sigma(r) u^c,
\end{equation}
as
\begin{equation}
F^{ab}_{;b} = - 4 \pi J^a,
\end{equation}
where, $\sigma(r) $ is the proper charge density of the
distribution. Hence the electromagnetic field tensor can be given as,

\begin{equation}
F_{ab}  =  E(r) (\delta_a^t \delta_b^r-\delta_a^r \delta_b^t),
\end{equation}
where $E(r)$ is the electric field.

Therefore, the energy-momentum tensors in the interior of the star can be expressed in the following form:

\begin{equation}
T^{0}_{0}= 8{\pi}{\rho}+\frac{1}{2}E^2
\end{equation}

\begin{equation}
T^{1}_{1}= 8{\pi}{p_r}-\frac{1}{2}E^2
\end{equation}

\begin{equation}
T^{2}_{2}= T^{3}_{3}= 8{\pi}{p_t}-\frac{1}{2}E^2
\end{equation}

The Einstein-Maxwell field equations with matter distribution as equation 3. are analogous with the transformations,

\begin{equation}
8{\pi}{\rho}+\frac{1}{2}E^2 = \frac{1}{r^2}[r(1-e^{-2\lambda})]'
\end{equation}

\begin{equation}
8{\pi}{p_r}-\frac{1}{2}E^2 = -\frac{1}{r^2}(1-e^{-2\lambda})+\frac{2\nu'}{r}e^{-2\lambda}
\end{equation}

\begin{equation}
8{\pi}{p_t}-\frac{1}{2}E^2 = e^{-2\lambda}(\nu''+{\nu'}^2+\frac{\nu'}{r}-{\nu}'{\lambda'}-\frac{\lambda'}{r})
\end{equation}

and
\begin{equation}
(r^{2}E)'=4{\pi}{r^2}{\sigma}{e^{\lambda}}
\end{equation}

where a `$\prime$' denotes differentiation with respect to the
radial parameter $r$. When E=0, the Einstein-Maxwell system given
above reduces to the uncharged Einstein system. The equation (15)
yields the expressed for E in the form

\begin{equation}
E(r)= \frac{4\pi}{r^2} \int_0^r
{x^2}{\sigma(x)}e^{\lambda(x)} dx= \frac{q(r)}{r^2}
\end{equation}

where q(r) is total charge of the sphere under consideration and $\sigma(r)$ is the
proper charge density.

The mass of a star in an uncharged system is generally defined by,
\begin{equation}
M(r)=4{\pi} \int_0^{r} {\rho(x)}{x^2} dx 
\end{equation}
Here $R$ is taken as the radius of our star model.

The equation of state is considered as,
\begin{equation}
p_r=\omega{\rho}
\end{equation}

\section{A particular class of solutions:}
We consider the the electric field intensity as
\begin{equation}
E^2(r)= \frac{4a^2 r^2}{(1+2a{r^2})^2}
\end{equation}

We observe that the function is regular if $a > 0$
A similar form of E can be used as shown by Tikekar et al.,   \cite{q}, Komathiraj et al.,   \cite{r} and Islam S. et al.,   \cite{s}.
We consider the gravitational potential
Z(r)   \cite{t} as,

\begin{equation}
Z(r)= \frac{(1+ar^2)(1-br^2)}{(1+2ar^2)}
\end{equation}

where a and b are real constants. Hence from above we observe,
\begin{equation}
e^{-2\lambda(r)}= \frac{(1+ar^2)(1-br^2)}{(1+2ar^2)}
\end{equation}
which on solving we get

\begin{equation}
\lambda'(r)= \frac{2ar}{(1+2a{r^2})}+\frac{br}{(1-br^2)}-\frac{ar}{(1+ar^2)}
\end{equation}
and
\begin{equation}
\lambda(r)=\frac{1}{2}[log(1+2a{r^2})-log(1+a{r^2})-log(1-b{r^2})]
\end{equation}

Using equations, (12), (19) and (21) we obtain
\begin{equation}
{\rho}= \frac{3(a+b)+ab r^2(6a r^2+7)}{8{\pi}(1+2a{r^2})^2}
\end{equation}

\begin{figure}[htbp]
\centering
\includegraphics[scale=.15]{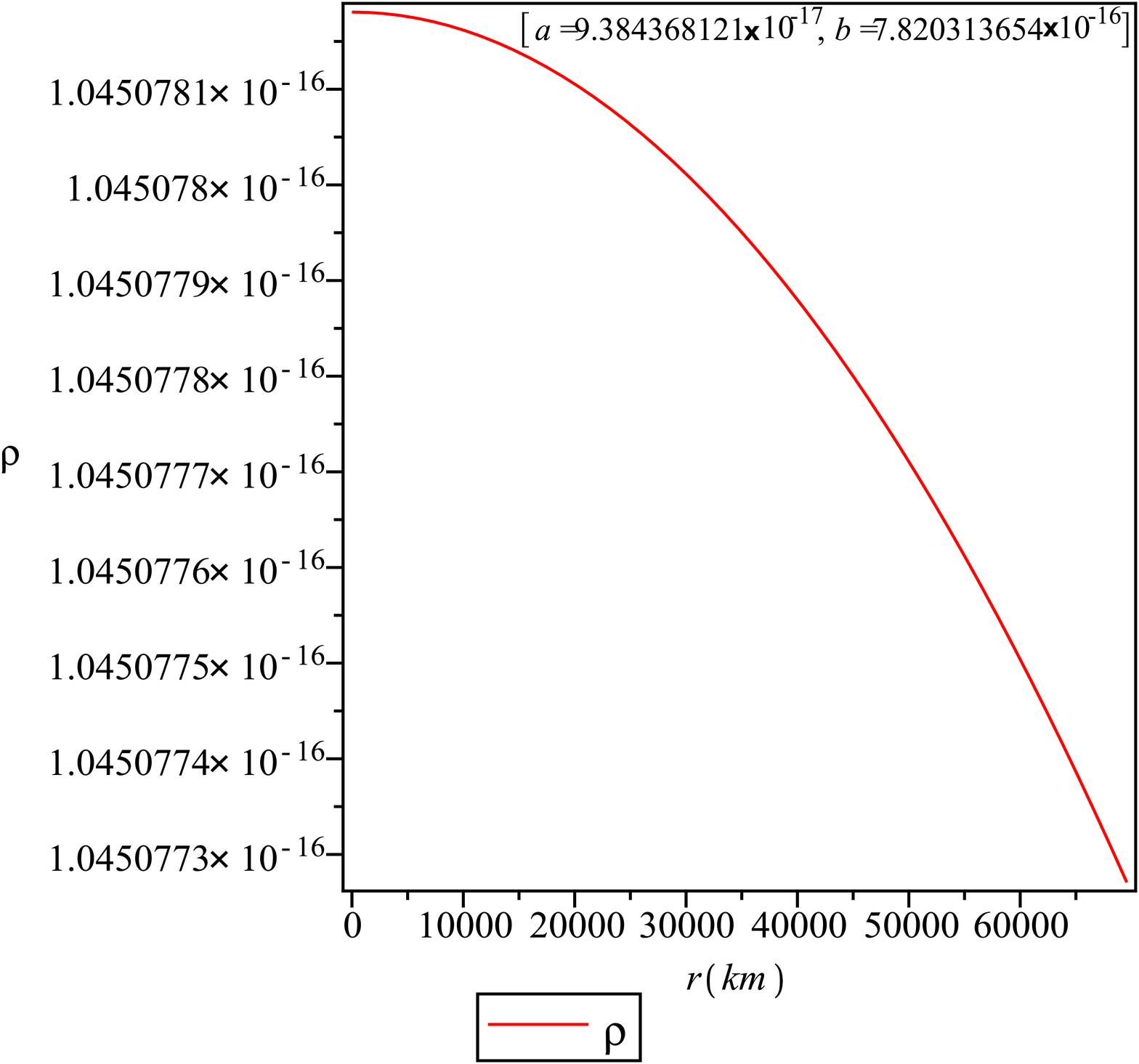}
\caption{The density parameter $\rho$ is shown against $r$, a and b having unit $km^{-2}$}
\end{figure}

Also for a positive density we must have,
\begin{equation}
-\frac{a}{1+\frac{7}{3}a r^2+2a r^4} < b,
\end{equation}

It is clearly evident from figure 1. that the density decreases gradually from the centre where it is
maximum and at the surface of the star of radius R, it becomes minimum. The variation is very small for the considered range of parameter. Hence the star is of almost uniform density having a value of
$1.04926~\times~10^{5}$~$\rm kg/m^{3}$.
[We have assumed that the radius of the star is $0.1~\times~R_{\odot}=69570 ~\rm km$].

Equations, (18) and (24) above yield
\begin{equation}
{p_r}= \omega \frac{3(a+b)+ab r^2(6a r^2+7)}{8{\pi}(1+2a{r^2})^2}
\end{equation}

Using equations, (14), (19) and (21), we get the value of the other parameter taking the constant of integration as zero without any loss of generality as,

\begin{eqnarray}
\nu(r)&=&\frac{(3{\omega}a+{\omega}b+a)}{2(2a+b)}log(1+2a{r^2})\nonumber \\
      & &-\frac{(3{\omega}a+2{\omega}b+a)}{4(a+b)}log(1+ar^2)\nonumber \\
      & &-\frac{(10{\omega}ab+3{\omega}{b^2}+6{\omega}{a^2}+2{a^2}+4ab+b^2)}{4(a+b)(2a+b)}\nonumber \\
      & &\times log(1-b{r^2})
\end{eqnarray}

Hence the following relation is evident,

\begin{align}
&& e^{2\nu(r)}=(1+2a{r^2})^\frac{(3{\omega}a+{\omega}b+a)}{2(2a+b)}\nonumber \\
         && \times (1+ar^2)^\frac{-(3{\omega}a+2{\omega}b+a)}{4(a+b)}\nonumber \\
           && \times (1-b{r^2})^\frac{-(10{\omega}ab+3{\omega}{b^2}+6{\omega}{a^2}+2{a^2}+4ab+b^2)}{4(a+b)(2a+b)}
\end{align}

\newpage

We also plot the metric potentials as below,
\begin{figure}[htbp]
\centering
\includegraphics[scale=.15]{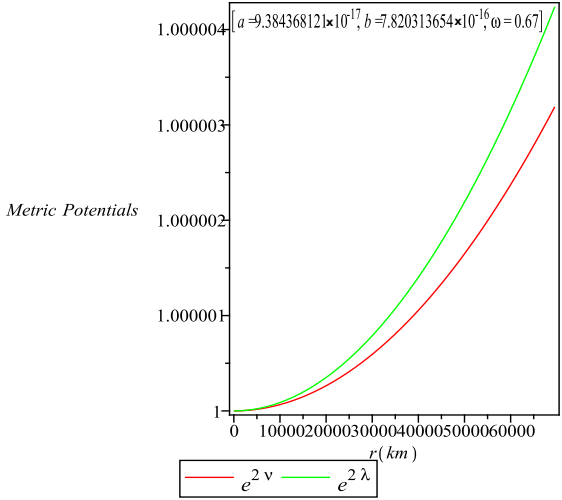}
\caption{The metric potentials $e^{2\nu}$ and $e^{2\lambda}$ are shown against $r$.}
\end{figure}

From the above figure we observe that both the metric potentials vanish at the centre, also
both $e^{2\nu}$ and $e^{2\lambda}$ increase with the increase in radius of the object.

From equation, (14) using (19), (21) and (26) we obtain,
\begin{eqnarray}
{p_t}&=&\frac{a(3{\omega}a+{\omega}b+a)(1+a{r^2})}{4{\pi}(2a+b)(1+2ar^2)^3}\nonumber \\
     & &\times (1-b{r^2})(1-2a{r^2})\nonumber \\
     & &+\frac{b(10{\omega}ab+3{\omega}b^2+6{\omega}a^2+2a^2+4ab+b^2)}{16{\pi}(a+b)(2a+b)(1+2a{r^2})(1-br^2)}\nonumber \\
     & &\times (1+a{r^2})(1+b{r^2})\nonumber \\
     & &-\frac{a(3{\omega}a+2{\omega}b+a)(1-a{r^2})(1-b{r^2})}{16{\pi}(a+b)(1+a{r^2})(1+2a{r^2})}\nonumber \\
     & &+\frac{{a^2(3{\omega}a+{\omega}b+a)^2}r^2(1+ar^2)(1-br^2)}{2{\pi}{(2a+b)^2}(1+2ar^2)^3}\nonumber \\
     & &+\frac{{a^2(3{\omega}a+2{\omega}b+a)^2}r^2(1-br^2)}{32{\pi}(a+b)^2(1+ar^2)(1+2ar^2)}\nonumber \\
     & &+\frac{{b^2(10{\omega}ab+3{\omega}b^2+6{\omega}a^2+2a^2+4ab+b^2)^2}}{32{\pi}(a+b)^2{(2a+b)^2}(1-br^2)(1+2ar^2)}\nonumber \\
     & &\times r^2(1+ar^2)\nonumber \\
     & &-\frac{a^2(3{\omega}a+{\omega}b+a)(3{\omega}a+2{\omega}b+a)}{4{\pi}(2a+b)(a+b)(1+2ar^2)^2}\nonumber \\
     & &\times r^2(1-br^2)\nonumber \\
     & &-\frac{ab(3{\omega}a+2{\omega}b+a)r^2}{16{\pi}{(a+b)^2}(2a+b)(1+2ar^2)}\nonumber \\
     & &\times (10{\omega}ab+3{\omega}b^2+6{\omega}a^2+2a^2+4ab+b^2)\nonumber \\
     & &+\frac{ab(3{\omega}a+{\omega}b+a)r^2(1+ar^2)}{4{\pi}(a+b){(2a+b)^2}(1+2ar^2)^2}\nonumber \\
     & &\times (10{\omega}ab+3{\omega}b^2+6{\omega}a^2+2a^2+4ab+b^2)\nonumber \\
     & &+\frac{a(3{\omega}a+{\omega}b+a)(1+ar^2)(1-br^2)}{4{\pi}(2a+b)(1+2ar^2)^2}\nonumber \\
     & &-\frac{a(3{\omega}a+2{\omega}b+a)(1-br^2)}{16{\pi}(a+b)(1+2ar^2)}\nonumber \\
     & &+\frac{b(10{\omega}ab+3{\omega}b^2+6{\omega}a^2+2a^2+4ab+b^2)(1+ar^2)}{16{\pi}(a+b)(2a+b)(1+2a{r^2})}\nonumber \\
     & &-[\frac{a(3{\omega}a+{\omega}b+a)r(1+ar^2)(1-br^2)}{4{\pi}(2a+b)(1+2ar^2)^2}\nonumber \\
     & &-\frac{a(3{\omega}a+2{\omega}b+a)r(1-br^2)}{16{\pi}(a+b)(1+2ar^2)}\nonumber \\
     & &+\frac{b(10{\omega}ab+3{\omega}b^2+6{\omega}a^2+2a^2+4ab+b^2)r(1+ar^2)}{16{\pi}(a+b)(2a+b)(1+2a{r^2})}]\nonumber \\
     & &\times [\frac{2ar}{(1+2ar^2)}+\frac{br}{(1-br^2)}-\frac{ar}{(1+ar^2)}]\nonumber \\
     & &-\frac{a(1+a{r^2})(1-b{r^2})}{4{\pi}(1+2ar^2)^2}-\frac{b(1+a{r^2})}{8{\pi}(1+2ar^2)}\nonumber \\
     & &+\frac{a(1-b{r^2})}{8{\pi}(1+2ar^2)}+\frac{a^2 r^2}{4{\pi}(1+2a{r^2})^2} 
\end{eqnarray}

\begin{figure}[htbp]
\centering
\includegraphics[scale=.13]{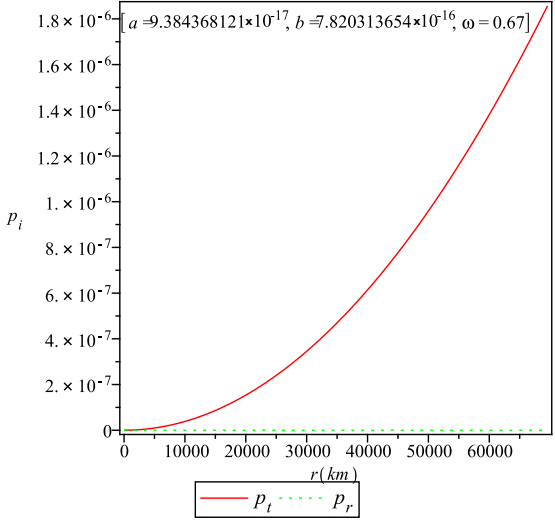}
\caption{The variation of radial and tangential pressure is plotted against $r$.}
\end{figure}

The radial pressure at the centre is equivalent to the tangential pressure at
the centre and found to be $8.49801~\times~10^{21}$ newtons~per~square~meter.
However the radial pressure decreases marginally with increase in distance from the
centre and at a distance 69570 km from the centre has the value
of $8.498~\times~10^{21}$ newtons~per~square~meter. The tangential pressure increases with distance and has a value
$2.253~\times~10^{32}$ newtons~per~square~meter at the surface of the star. 
The difference between the tangential and radial pressure gives a measure of pressure anisotropy.
It is positive throughout the interior that is $p_t > p_r$. We can thus conclude that the anisotropic pressure is repulsive in nature.
Hence there is an external thrust acting on the body.

The pressure anisotropy is represented by fig.4,
\begin{figure}[htbp]
\centering
\includegraphics[scale=.4]{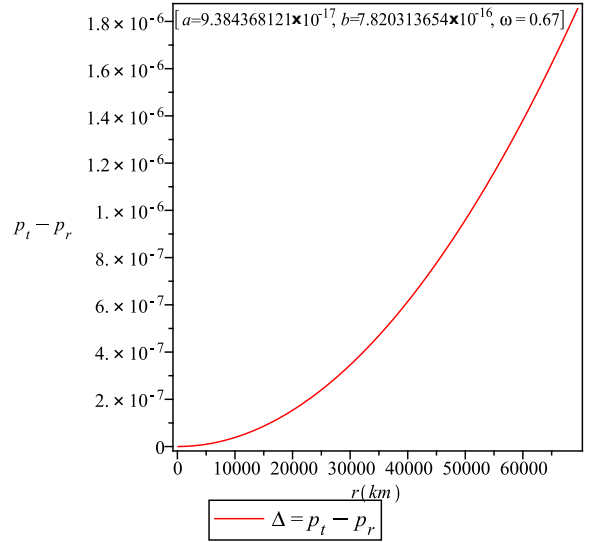}
\caption{The pressure anisotropy $\Delta$ is shown against r.}
\end{figure}

\section{Stability analysis:}
To discuss the stability of the star under consideration we
have considered the Hererra's approach   \cite{u}. which is known as the
concept of cracking (or overturning). This theorem states  that
the region for which $v_{st}^2 -  v_{sr}^2<0 $ is a stable region
and the region for which $v_{st}^2 - v_{sr}^2>0 $ is an  unstable
$v_{sr}^2$ [Here $\frac{dp_t}{d \rho}=v_{st}^2$ and $\frac{dp_r}{d \rho}=v_{sr}^2$;
$v_{st}$ and $v_{sr}$ stand for the tangential and radial velocity of sound respectively.]
We note from the curve profile (fig.5) for $v_{st}^2 -
v_{sr}^2 $ that it is negative throughout, hence the star is potentially
stable throughout the interior region.

\begin{figure}[htbp]
\includegraphics[scale=.15]{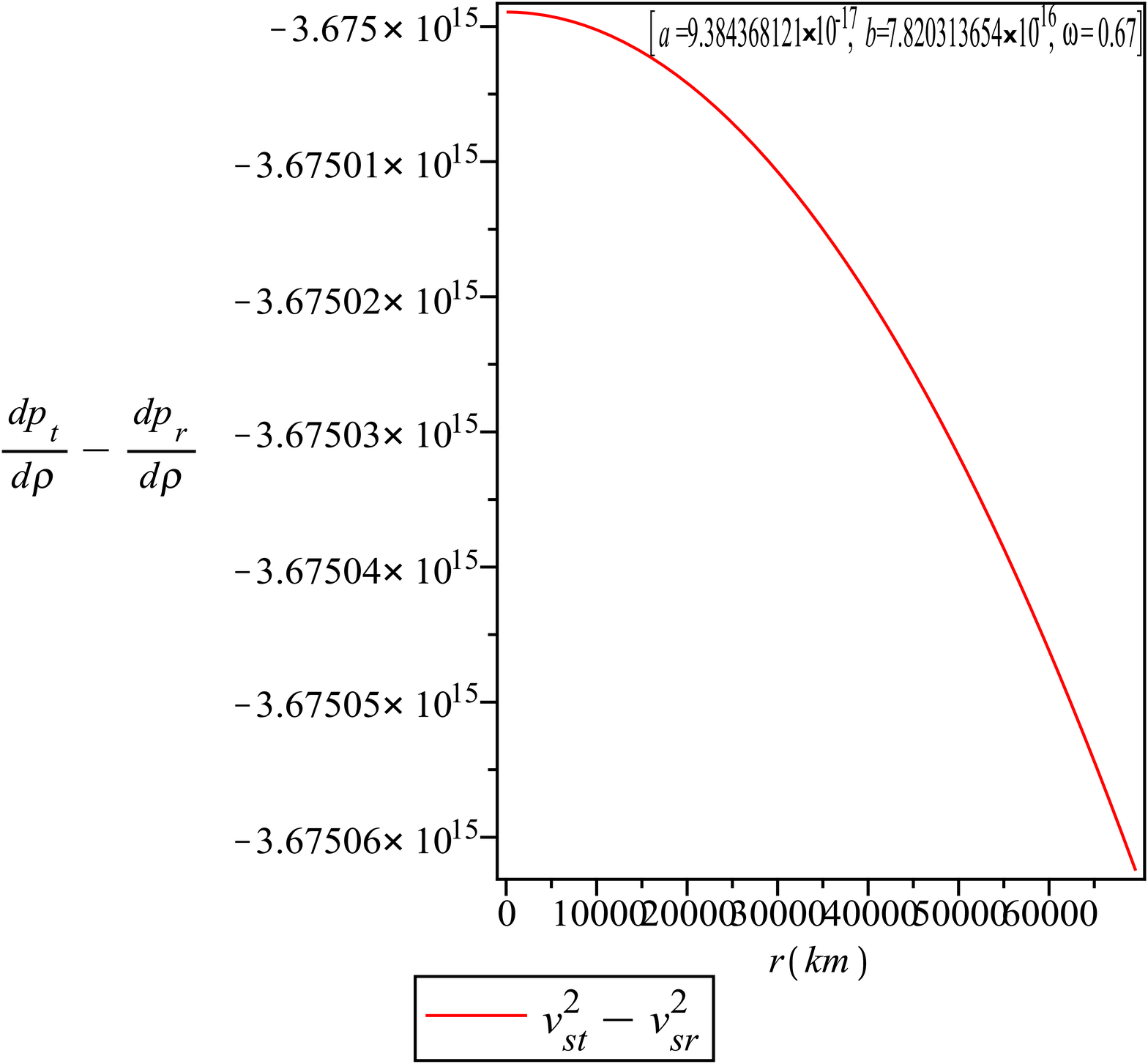}
\caption{The variation of $ v_{st}^2 - v_{sr}^2 $ is shown against
$r$.} 
\end{figure}

\section{Electric Field and proper charge density:}
~~~~~~~~~~The expressions for Electric field is
\begin{equation}
E(r)=\frac{2ar}{(1+2a{r^2})}
\end{equation}

Also the charge is given by,
\begin{equation}
q(r)=\frac{2a r^3}{(1+2a{r^2})}
\end{equation}

Now, at the centre and at the boundary of the star, we have
\begin{equation}
 E_{(r=0)}=0
\end{equation}

\begin{eqnarray}
E_{(r=R)}&=&\frac{2a R}{(1+2a{R}^2)} \nonumber\\
         &=& 1.36344 \times 10^{13}~\rm newtons~p.c
\end{eqnarray}

The corresponding values of charge are  given by,
\begin{equation}
q_{(r=0)}=0
\end{equation}

\begin{eqnarray}
q_{(r=R)}&=&\frac{2a {R}^3}{(1+2a{R}^2)} \nonumber\\
         &=& 7.34538 \times 10^{18}~ \rm coulomb
\end{eqnarray}

We observe that both the electric field intensity and charge vanishes at the centre
and both increases radially outward, which are illustrated in fig.6 and fig.7

\begin{figure}[htbp]
\centering
\includegraphics[scale=.15]{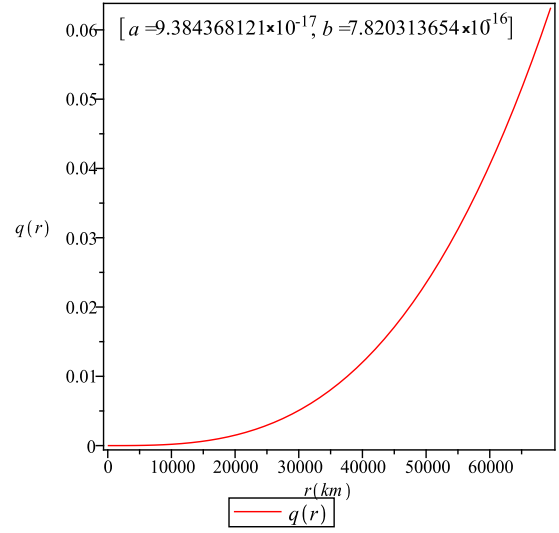}
\caption{The variation of the charge $q(r)$ is
shown against $r$.}
\end{figure}

\begin{figure}[htbp]
\centering
\includegraphics[scale=.16]{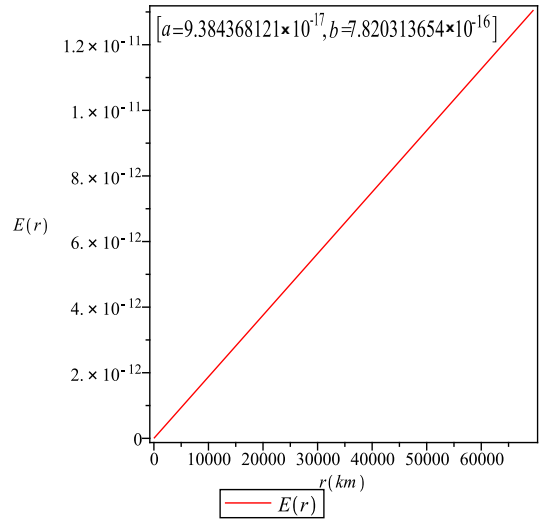}
\caption{The variation of electric field is shown against $r$.}
\end{figure}

The proper charge density, using equation.(15) is given by,
\begin{equation}
\sigma(r)=\frac{a (3+2a r^2)(1+a{r^2})^{\frac{1}{2}}(1-b{r^2})^{\frac{1}{2}}}{2{\pi} (1+2ar^2)^\frac{5}{2}} 
\end{equation}

It is evaluated to be $6.0398~\times~10^{4}$~coulomb per $\rm metre^{3}$ at the centre
and $6.03978~\times~10^{4}$~coulomb per $\rm metre^{3}$ at the surface
(69570 km from the centre) of the star.

The variation of the proper charge density with the radius is shown in fig.8.

\begin{figure}[htbp]
\centering
\includegraphics[scale=.15]{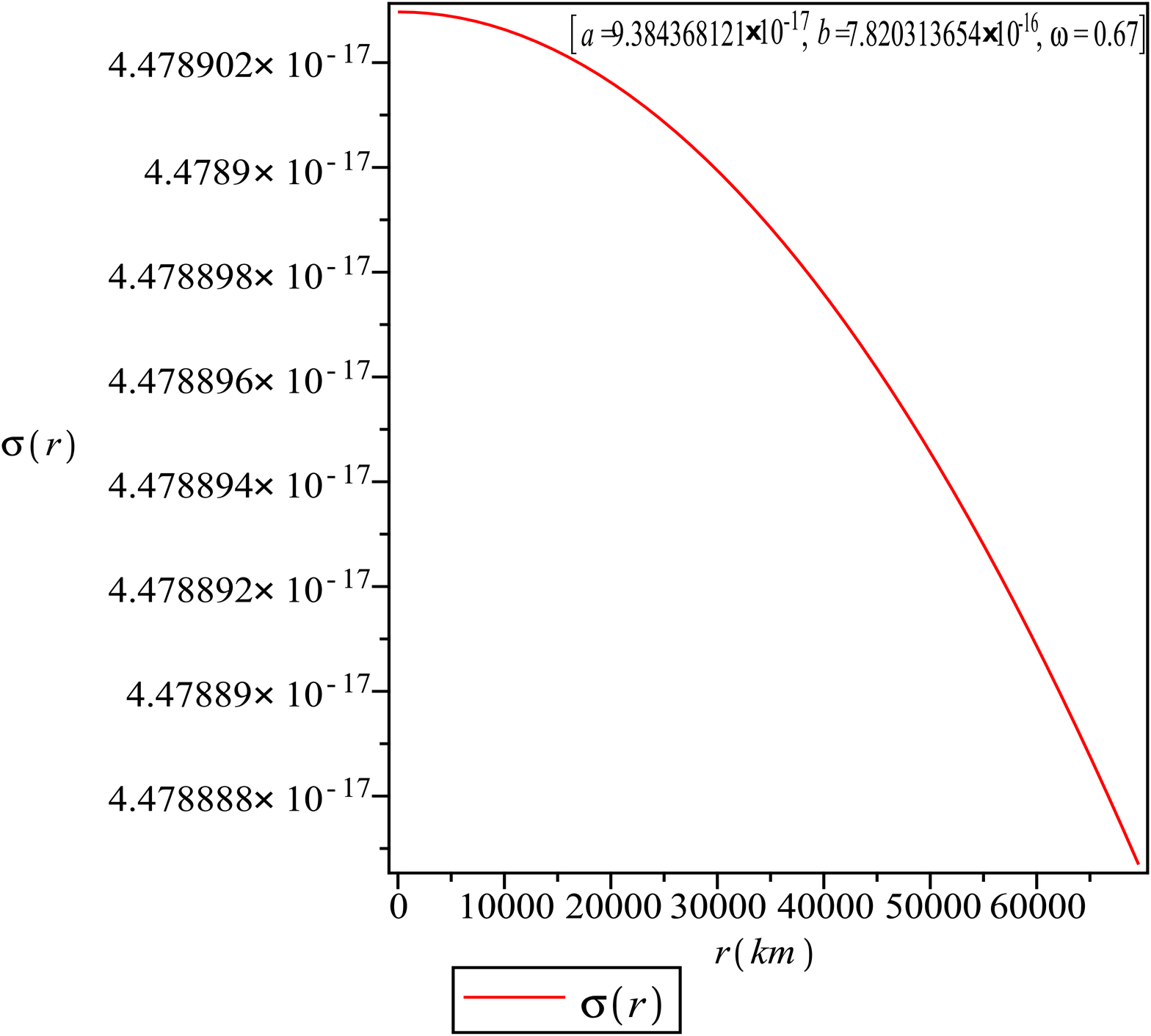}
\caption{The variation of the proper charge density $\sigma(r)$ is plotted
against $r$.}
\end{figure}

\newpage
\section{Mass radius relation:}
Now, we calculate the effective mass which is given by,

\begin{eqnarray}
M_{eff}&=&4\pi \int_0^r \left[\rho+\frac{E^2}{16\pi}\right]r^2 dr \nonumber \\
       &=&\frac{(a+b+ab r^2) r^3}{2(1+2a r^2)}
\end{eqnarray}

We observe that, since the effective mass is always positive,

\begin{equation}
b > -\frac{a}{(1+a r^2)}
\end{equation}

\begin{figure}[htbp]
\centering
\includegraphics[scale=.15]{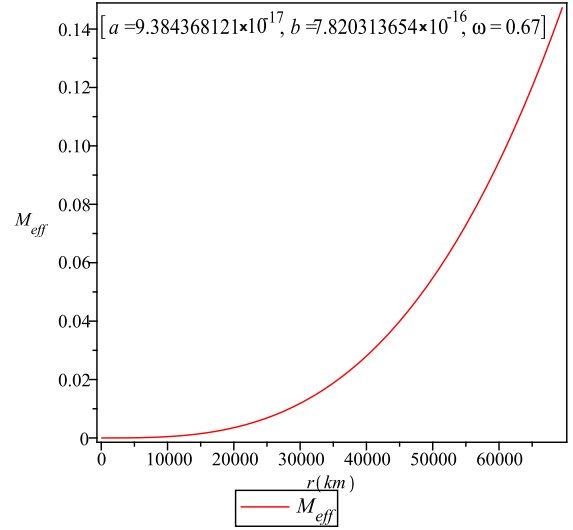}
\caption{The effective mass is plotted against $r$.}
\end{figure}

\begin{equation}
\frac{2M_{eff}}{r}= \frac{(a+b+ab r^2) r^2}{(1+2a r^2)}
\end{equation}

The condition for a star not to collapse gravitationally, in general, is given by

\begin{equation}
\frac{2M_{eff}}{r}<1
\end{equation}
The above condition can be taken as a rough estimate for a charged\cite{w} anisotropic star.
Hence the above equation implies that
\begin{equation}
b < \frac{1}{r^2}
\end{equation}

Hence the above equations imply,
\begin{equation}
-\frac{a}{(1+a r^2)} < b < \frac{1}{r^2};~~~~~~~~ a>0 
\end{equation}

Now since,
\begin{equation}
-\frac{a}{(1+a r^2)} < -\frac{a}{1+\frac{7}{3}a r^2+2a^2 r^4} 
\end{equation}

Hence from equations, (40), (43), (44) and (45) we get the range of parameter b as,

\begin{equation}
-\frac{a}{1+\frac{7}{3}a r^2+2a^2 r^4}< b < \frac{1}{r^2};~~~~~~~~ a>0 
\end{equation}

At $r=\frac{1}{\sqrt{2a}}$, where the electric field is maximum we observe that,
\begin{equation}
-\frac{3a}{8} < b < 2a;~~~~~~~~ a>0
\end{equation}

Also for the values of the parameter $a=9.384368 \times {10}^{-17}~\rm km^{-2}$, $b=7.820314 \times {10}^{-16}~ \rm km^{-2}$ and $r=69570 ~\rm km$,
we observe that the condition for a charged star is satisfied as,

\begin{equation}
\frac{2M_{eff}}{r}=2.87399 \times {10}^{-6} < 1
\end{equation}

The compactness of the star is obtained as,

\begin{eqnarray}
u &=&\frac{M_{eff}}{r} \nonumber\\
  &=&\frac{(a+b+ab r^2) r^2}{2(1+2a r^2)},
\end{eqnarray}

which is illustrated graphically in fig.10.

\begin{figure}[htbp]
\centering
\includegraphics[scale=.18]{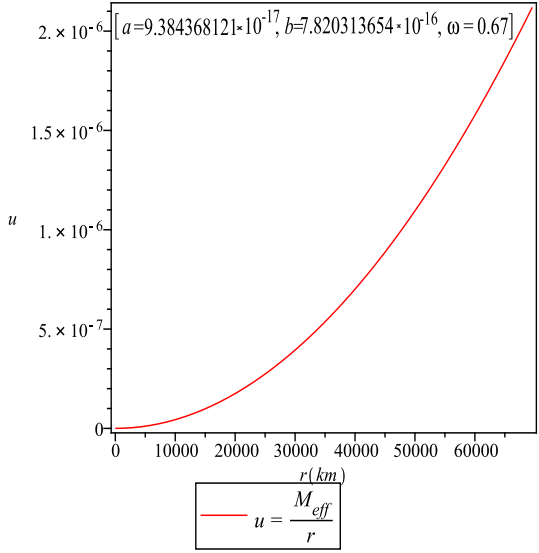}
\caption{The compactness of the star is shown against its radius $r$.}
\end{figure}

The surface red shift function is obtained as,
\begin{eqnarray}
Z_{s}&=&[1-(2u)]^{-\frac{1}{2}}-1 \nonumber\\
     &=&[\frac{(1+2a r^2)}{1+(a-b) r^2 - ab r^4}]^{\frac{1}{2}}-1,
\end{eqnarray}

The variation of the surface red shift $Z_s$ is shown in fig.11.
\begin{figure}[htbp]
\centering
\includegraphics[scale=.15]{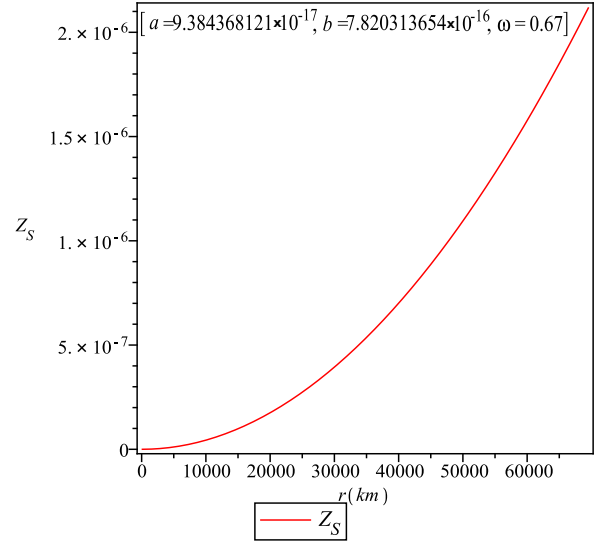}
\caption{The red-shift function of the star is shown against its radius $r$.}
\end{figure}

What happens after a low-mass star ceases to produce energy through fusion has not yet been directly observed; the universe is around 13.8 billion years old,
which is less time (by several orders of magnitude, in some cases) than it takes for fusion to cease in such type of stars. We calculate the surface red shift
as $Z_s= 2.12 \times {10}^{-6}$. 

We estimate the effective mass of the star as
$0.1M_\odot$. For all these estimations, we have taken the radius
of the star as $0.1R_\odot$ and values of the constants as
$a=9.384368 \times {10}^{-17}~\rm km^{-2}$, $b=7.820314 \times {10}^{-16}~\rm km^{-2}$ and $\omega=0.67$.

\section{Energy conditions:}

All the energy conditions, namely, null energy condition (NEC),
weak energy condition (WEC) and  strong energy condition (SEC)
  are satisfied not only at the centre ($r=0$) but throughout the interior region (fig.12):
\begin{equation}
~~~~~~~~~~~~~~~~\rho+\frac{E^2}{16{\pi}}{\geq} 0,
\end{equation}
\begin{equation}
~~~~~~~~~~~~~~~~\rho+p_r{\geq} 0
\end{equation}
\begin{equation}
~~~~~~~~~~~~~~~~\rho+p_t+\frac{E^2}{8{\pi}}{\geq} 0
\end{equation}
\begin{equation}
~~~~~~~~~~~~~~~~\rho+p_r+2p_t+\frac{E^2}{8{\pi}}{\geq} 0
\end{equation}

\begin{figure}[htbp]
\centering
\includegraphics[scale=.14]{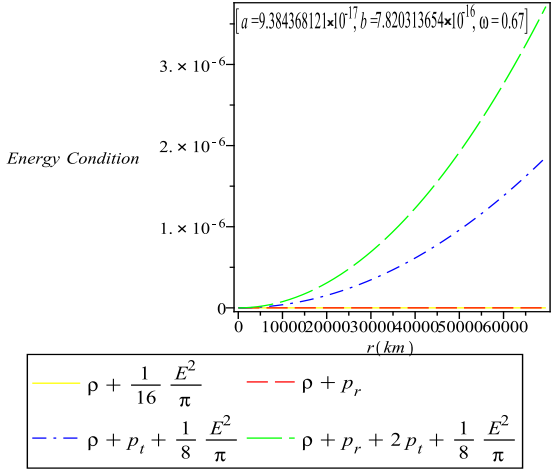}
\caption{The energy condition of the system has been plotted against $r$.}
\end{figure}

\section{Generalized TOV equations}
Now we can write the
generalized Tolman-Oppenheimer-Volkoff (TOV) equations, 
which gets the form
\begin{equation}
-\frac{M_G\left(\rho+p_r\right)}{r^2}e^{\lambda-\nu}-\frac{dp_r}{dr}
+\sigma\frac{q}{r^2}e^{\lambda}
+\frac{2}{r}\left(p_t-p_r\right) = 0. 
\end{equation}
Here, $M_G$ is the effective gravitational mass given by
\begin{equation}
M_G(r)= r^2e^{\nu-\lambda}\nu^{\prime}. 
\end{equation}
The above equation describes the equilibrium condition for the
charged star subject to the following forces such as
gravitational ($F_g$), hydrostatic ($F_h$), electric ($F_e$) and
anisotropic stress ($F_a$) so that
\begin{equation}
F_g+ F_h+ F_e+ F_a=0
\end{equation}
where,
\begin{eqnarray}
F_g &=& -\nu'\left(\rho+p_r\right)  \\ 
F_h &=&  -\frac{dp_r}{dr}  \\ 
F_e &=& \sigma E e^{\lambda}   \\ 
F_a &=& \frac{2}{r}\left(p_t-p_r\right), 
\end{eqnarray}

In the figures (13-16), we have shown the profiles of these forces at the
interior of the star. The equilibrium stage is achieved under the combined effects of
these forces. The gravitational force is balanced by the anisotropic, hydrostatic and
the electrical force thereby making the system balanced.

\begin{figure}[htbp]
\centering
\includegraphics[scale=.15]{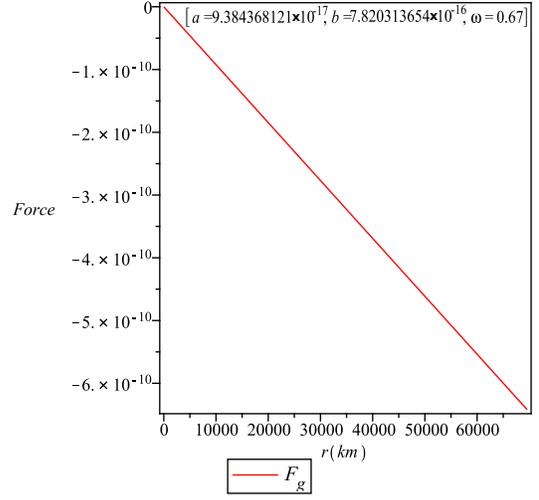} \caption{Gravitational
force acting on interior of the star in static equilibrium.}
\end{figure}

\begin{figure}[htbp]
\centering
\includegraphics[scale=.12]{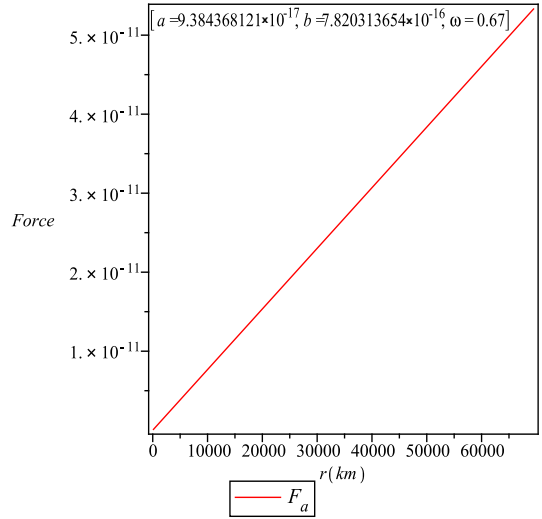} \caption{Anisotropic
force acting on interior of the star in static equilibrium.}
\end{figure}

\begin{figure}[htbp]
\centering
\includegraphics[scale=.14]{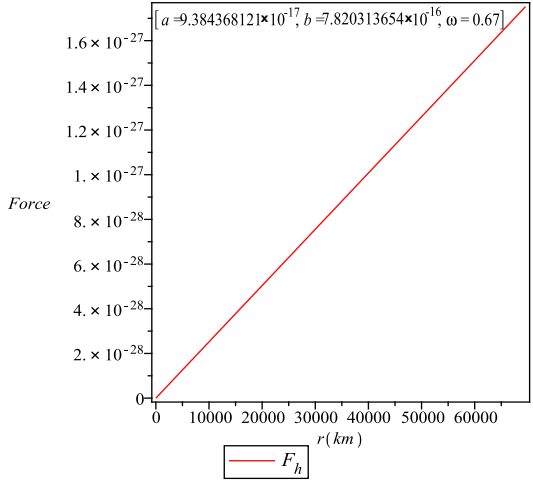} \caption{Hydrostatic
force acting on interior of the star in static equilibrium.}
\end{figure}

\begin{figure}[htbp]
\centering
\includegraphics[scale=.13]{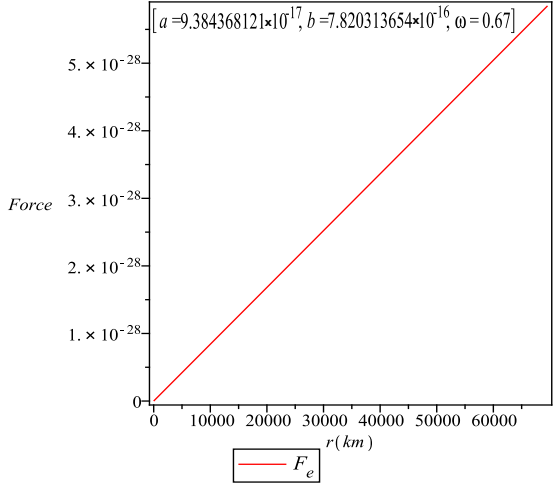} \caption{Electric
force acting on interior of the star in static equilibrium.}
\end{figure}

\newpage
\section{Gravitational potential}
The gravitational potential at the surface of the star is evaluated from equation.(20) as $Z(r)=3.48062 \times 10^{18}$~~Joules~per~kg. It has been shown graphically,

\begin{figure}[htbp]
\centering
\includegraphics[scale=.15]{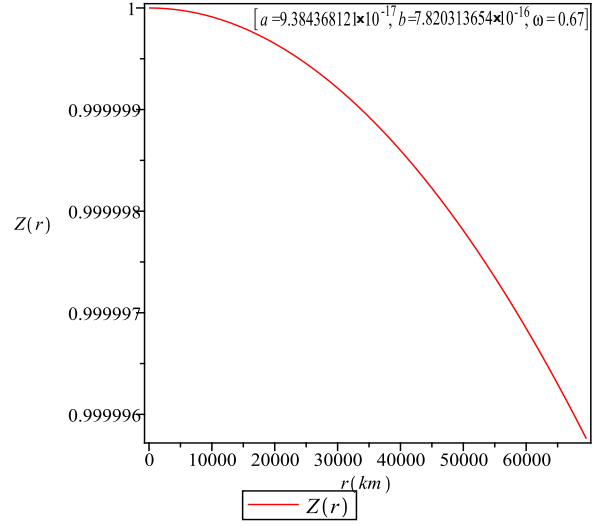} \caption{The gravitational potential $Z(r)$ is plotted against $r$.}
\end{figure}

We also express the total gravitational energy K(r) using equations, (20) and (39) as follows,

\begin{align*}
 K(r)=M_{eff}{\times}Z(r)
\end{align*}
\begin{align}
&&= \frac{(a+b+ab r^2) r^3}{2(1+2a r^2)}.\frac{(1+ar^2)(1-br^2)}{(1+2ar^2)} \nonumber\\\nonumber\\
&&= \frac{(a+b+ab r^2)(1+ar^2)(1-br^2) r^3}{2 (1+2ar^2)^2},
\end{align}

We have evaluated the total gravitational potential energy of the system to be $1.78962~\times~10^{46}$ Joules.
\newpage
\section{a-b Parameter Space}
 a and b are two parameters in this model, both of them having dimension $[L]^{-2}$. For simplicity through out the whole discussion, we use the dimension to be $\frac{1}{R_{\odot}^2}$ where $R_\odot$ is the solar radius. We simply denote $M_{eff}$ by $M$. In this section, we find the allowable region in a-b parameter space for physically possible stars according to our model. Inequality (44) gives the allowable region in a-b parameter space. Considering different radii of stars, one can find respective allowable regions in a-b parameter space.
 \begin{figure}[htbp]
\includegraphics[scale=0.42]{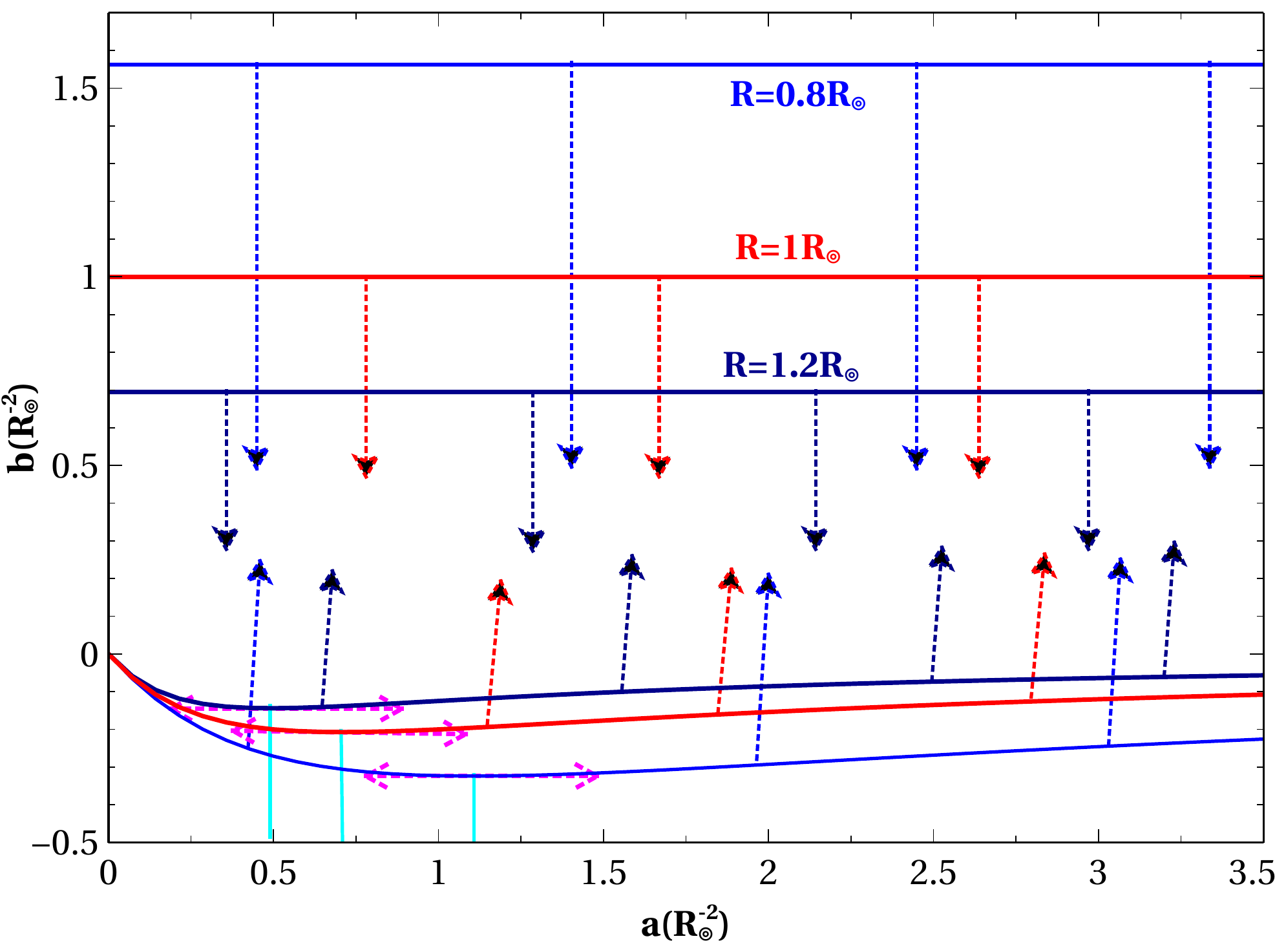}
 \caption{Allowable regions in a-b parameter space for different values of radius of star}
 \end{figure}

With the increase in radius, the allowable region in a-b parameter space shrinks towards $b=0$ line and it is obvious from the expression of inequality (44) too. It is also worth mentioning that $a\geq0$. The allowable region for a fixed radius represents all possible mass configuration of a star for that radius in our model. The upper limit of b in inequality (44) does not depend on a and is equal to the inverse square of the chosen radius. The curve corresponding to the lower limit of b has a minima at $a=\frac{1}{\sqrt{2}r^2}$ which is shown for three different values of stellar radius in Fig. 18. For $a=0$, lower limit of $b=0$. As $a\rightarrow\infty$, lower limit of  b  $\rightarrow 0-$.\\
The mass of a star is a function of r, a and b. When the stellar radius, r is chosen to be fixed, mass of the star is only function of a and b. Hence a single curve in a-b parameter space obeying equation(39) represents a star, i.e; any a and b chosen from that line represents same mass and same radius. This degeneracy is occurring due to the fact that the points from equal mass-radius line represents different possible charge configurations of that star according to the equation (31). As the charge of a star is not a directly measurable quantity, one can not take charge as input, hence one can not get rid of this degeneracy. Nevertheless one can find the behaviour of stellar mass with parameter a and b when the stellar radius is fixed, i.e; the allowable region in a-b parameter space is decided.
\begin{figure}[htbp]
\includegraphics[scale=0.132]{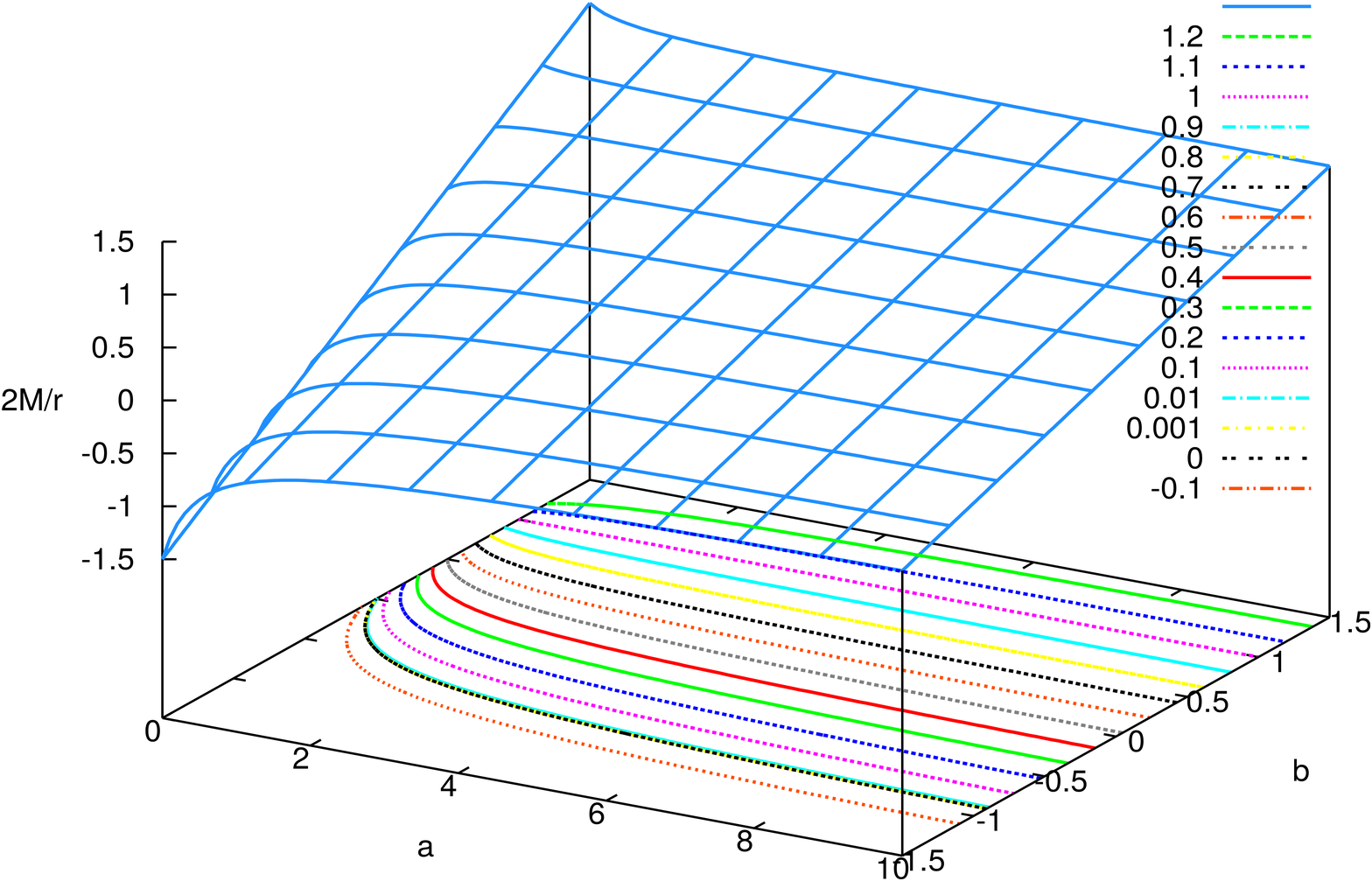}
\caption{Contour plot of $\frac{2M}{r}$ with a and b for stellar radius $1~R_\odot$}
\end{figure}
In Fig. 19 several values (both physical and unphysical) of $\frac{2M}{r}$ is shown and for each value of $\frac{2M}{r}$ a contour is drawn in a-b plane.
\begin{figure}[htbp]
\includegraphics[height=7.1 cm, width=8.417 cm]{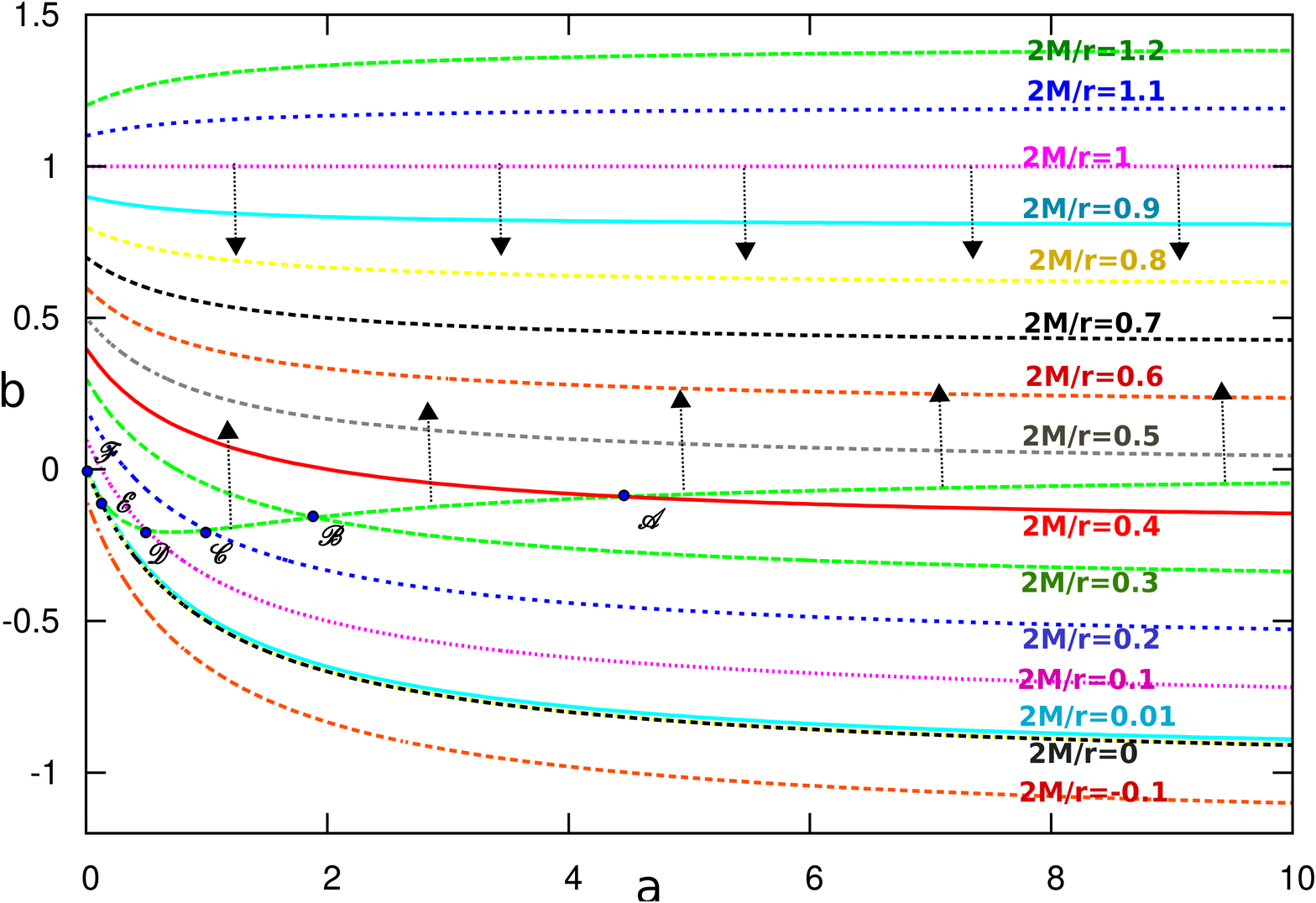}
\caption{Projection of Fig. 19 on a-b plane is plotted with the allowable region for $r=1~R_\odot$.}
\end{figure}

Equation (39) implies that $\frac{2M}{r}$ increases linearly with b and for $a\geq0$, the surface corresponding to $\frac{2M}{r}$ has no extrema.\\
Two curves with arrows in Fig. 22 represent the lower and upper bounds of the allowable region in a-b parameter space. The equal mass-radius curves having negative $\frac{2M}{r}$ are outside the feasible region. One can reach this argument using equation(41)-(43) too. The curves corresponding to $\frac{2M}{r}\geq 1$ are also outside our feasible region in a-b parameter space. Equation(41) demonstrates the fact. Among the curves in the feasible region of a-b parameter space, some of them (e.g. curves corresponding to $\frac{2M}{r}=0.4,0.3,0.2 $ etc)  intersect the lower bound line, i.e; for those curves the whole length is not permissible. \\
From equation (39)
\begin{equation}
b=\frac{(1+2ar^2)\frac{2M}{r}-ar^2}{(1+ar^2)r^2}
\end{equation}
For fixed mass and fixed radius, equation (61) represents the equal mass-radius line on a-b parameter space. For $a=0$, $b=\frac{2M}{r^3}$. $b>0~\forall~a\geq0$ for $\frac{2M}{r}>0.5$. For $\frac{2M}{r}=0.5,~b\geq 0 $ and b approaches $0$ as a$\rightarrow\infty$.
\begin{equation}
\lim\limits_{a\rightarrow\infty}b=\frac{1}{r^2}\left(\frac{4M}{r}-1\right)
\end{equation}
b asymptotically approaches a positive finite value as a tends to infinity for $\frac{2M}{r}>0.5$ and a negative finite value for $\frac{2M}{r}<0.5$. It is also worth mentioning that this curve has no maxima or minima for $a\geq0$. For $\frac{2M}{r}=1$, the curve is a straight line parallel to a-axis. The curve corresponding to the lower limit of b is given by
\begin{equation}
b=-\frac{a}{1+\frac{7}{3}a r^2+2a^2 r^4}
\end{equation}
$b\leq0~\forall~a\geq0$ and $b=0$ at $a=0$. This curve approaches $0$ as $a\rightarrow\infty$. Hence the equal mass-radius curves corresponding to $\frac{2M}{r}\geq0.5$ do not intersect the curve corresponding to the lower limit of b. Hence one can find two class of stars within the feasible region of a-b parameter space categorized by $0.5\leq\frac{2M}{r}<1$ and $0<\frac{2M}{r}<0.5$. The curve corresponding to $\frac{2M}{r}=0$ intersect the line corresponding to the lower bound of b at $a=0,b=0$. All of the equal mass-radius lines corresponding to $0<\frac{2M}{r}<0.5$, have finite length in the feasible region of a-b parameter space. Equating equation(61) and equation(63) gives the intersections of equal mass-radius curve with the curve corresponding to the lower limit of b in a-b parameter space and it is given by a cubic equation in a as
\begin{equation}
6r^6(2n-1)a^3+4r^4(5n-1)a^2+13r^2na+3n=0
\end{equation}
where $n=\frac{2M}{r}$.\\\\
Clearly for $\frac{2M}{r}=n=0.5$, the equation is a quadratic equation in a. The roots of this quadratic equation are real and negative, hence the roots lie outside the feasible region in a-b parameter space.\\
For $n=0$, equation (64) has two solutions, i.e; $a=0$ and $a=-\frac{2}{3r^2}$. $a=0$ is a root with multiplicity $2$ with corresponding solution for $b=0$ and $a=-\frac{2}{3r^2}$ lie outside the feasible region in a-b parameter space.\\
For $n=0.2$, the equation is in depressed cubic form. We solve it analytically by Cardano's method. The cubic equation for $n=0.2$ is given by
\begin{equation}
18x^3-13x-3=0
\end{equation}
where $x=ar^2$.\\
Substituting
\begin{align*}
x=u+v
\end{align*}
and setting
\begin{align*}
& 18(u^3+v^3)=3\\
& (uv)^3=\left(\frac{13}{54}\right)^3
\end{align*}
\begin{equation}
u^3=\frac{1}{12}+\sqrt{\frac{1}{144}-\frac{1}{27}\left(\frac{13}{18}\right)^3}
\end{equation}
and
\begin{equation}
v^3=\frac{1}{12}-\sqrt{\frac{1}{144}-\frac{1}{27}\left(\frac{13}{18}\right)^3}
\end{equation}
The term inside the square root in the expressions of $u^3$ and $v^3$ is negative, hence we write
\begin{align*}
& u=Ae^{i\theta}\\
& v=Ae^{-i\theta}
\end{align*}
where $i=\sqrt{-1}$.\\
We find $A=0.490653378$ and $\theta$ has three possible values: $15.04346531^\circ$, $135.04346531^\circ$ and $255.04346531^\circ$. Thus three solutions of $x(=2Acos\theta): x1,~x2$ and $x3$ have values
\begin{align*}
& x1=0.947676593\\
& x2=-0.694414854\\
& x3=-0.253261738
\end{align*}
$\because a(=\frac{x}{r^2})>0$, only positive $x$ is considerable. For $r=1~R_\odot$, $a=0.947676593~R_\odot^{-2}$ which matches with numerics and with Fig. 20. One can find the corresponding value of  b  from equation (61) or equation (63), $b=-0.189254191~R_\odot^{-2}$.\\
Generally $\forall~ 0< n\leq 1(n\neq0.5)$, finding a solution by Cardano's method is little clumsy. Nevertheless, one can find the nature of the roots by using Descartes' rule of signs. i.e; counting the sign changes for the polynomial $P(a)(=6r^6(2n-1)a^3+4r^4(5n-1)a^2+13r^2na+3n)$ from equation (64) and $P(-a)$ to find the number of real positive roots and the number of real negative roots respectively. We find that for $n>0.5$, all of three roots are negative; for $0<n<0.5$, one root is positive and the other two roots are negative. This conclusion matches with Fig. 20 and with the numerics done in the following section.
\section{Stars from H-R diagram in a-b parameter space}
We consider several observed stars from Hertzsprung-Russell diagram   \cite{a2} and we see where do those stars fit in our model, i.e; in a-b parameter space. Using equation (39), lines for those stars in the feasible regions are drawn. We use mass and radius of those stars and find the equal mas-radius lines for them. Other physical properties of those stars are not considered, the main aim of this section is to find some idea about a-b parameter space for the real stars.
\begin{figure}[htbp]
\includegraphics[height=7.5 cm, width=8.417 cm]{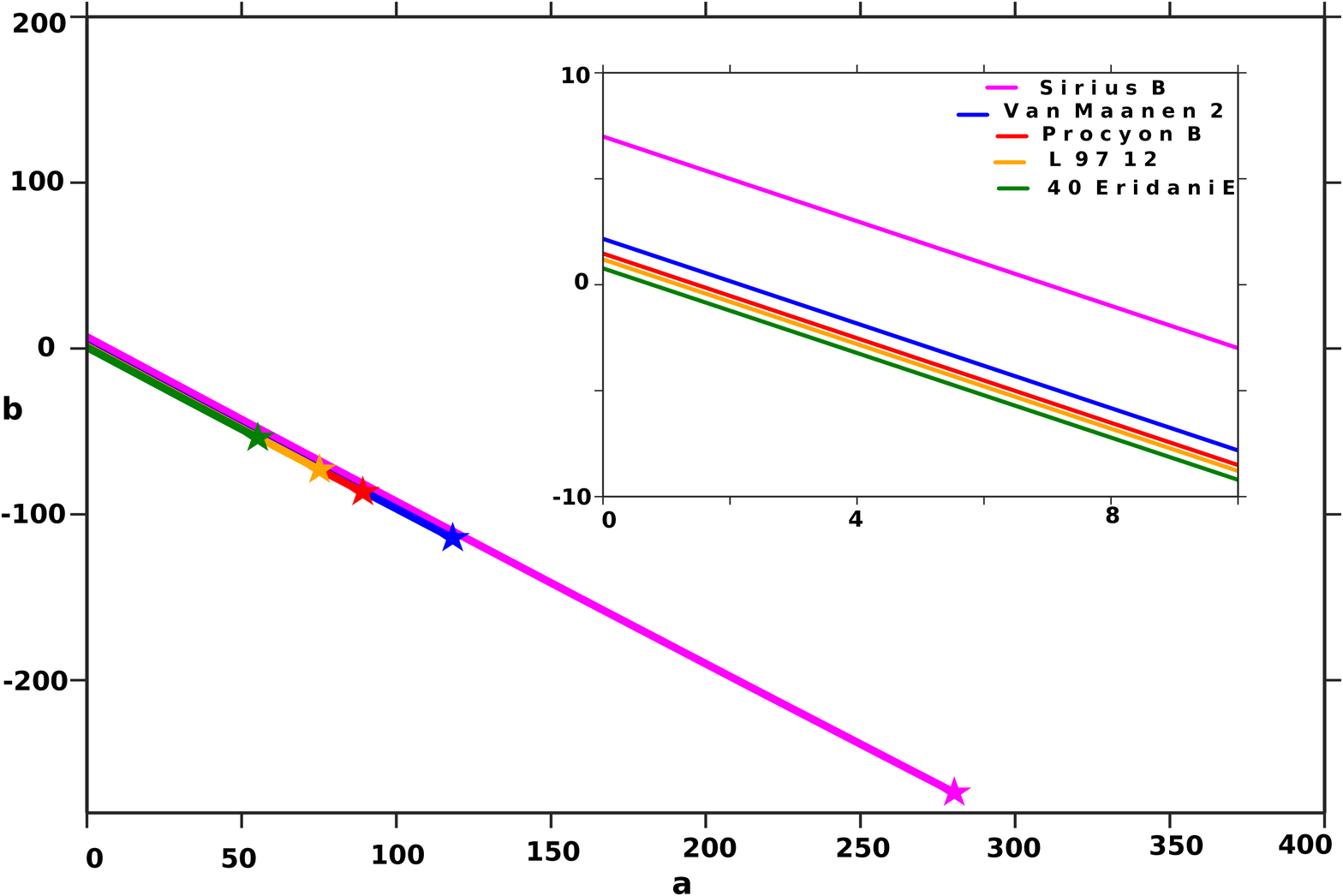}
\caption{White Dwarfs, the intersection points with the lower bound line of b are shown by '$\star$' s. All of the stars have $n<0.5$.}
\end{figure}

Temperature of white dwarfs \cite{a3}-\cite{a5}, having radius about 0.01 $R_{\odot}$, varies from 5000 to 30,000 K with a variation in luminosity nearly about $10^{-4}$ to $10^{-2}$$L_{\odot}$($L_{\odot}=$ Luminosity of the Sun= $3.846\times10^{26}$ W) in H-R diagram.
Fig. 21 shows that $a\sim[0,100]$ and $b\sim[-100,10]$ for the white dwarfs in consideration.\\
Main sequence is a distinctive continuous band of stars that appear in the H-R diagram (the Sun is in MS category) having stars of relatively smaller radii, cooler and less luminous than the Sun as well as stars, bigger, hotter and with a luminosity higher than the Sun. Among those MS stars, red dwarfs \cite{a6}-\cite{a9} have a cooler surface temperature than the Sun, typically around 3,500 K. Red dwarfs, relatively dimmer than the Sun, have radius nearly about 0.1 $R_\odot$ to 1 $R_\odot$.
Fig. 22 shows that $a\sim[0,0.1]$ and $b\sim[-0.1,10^{-4}]$ for the red dwarfs in consideration.
Fig. 23 shows that $a\sim[0,0.001]$ and $b\sim[-0.001,10^{-6}]$ for the considered MS stars \cite{a10}- other than red dwarfs.
\begin{figure}[h!]
\includegraphics[height=7.5 cm, width=8.417 cm]{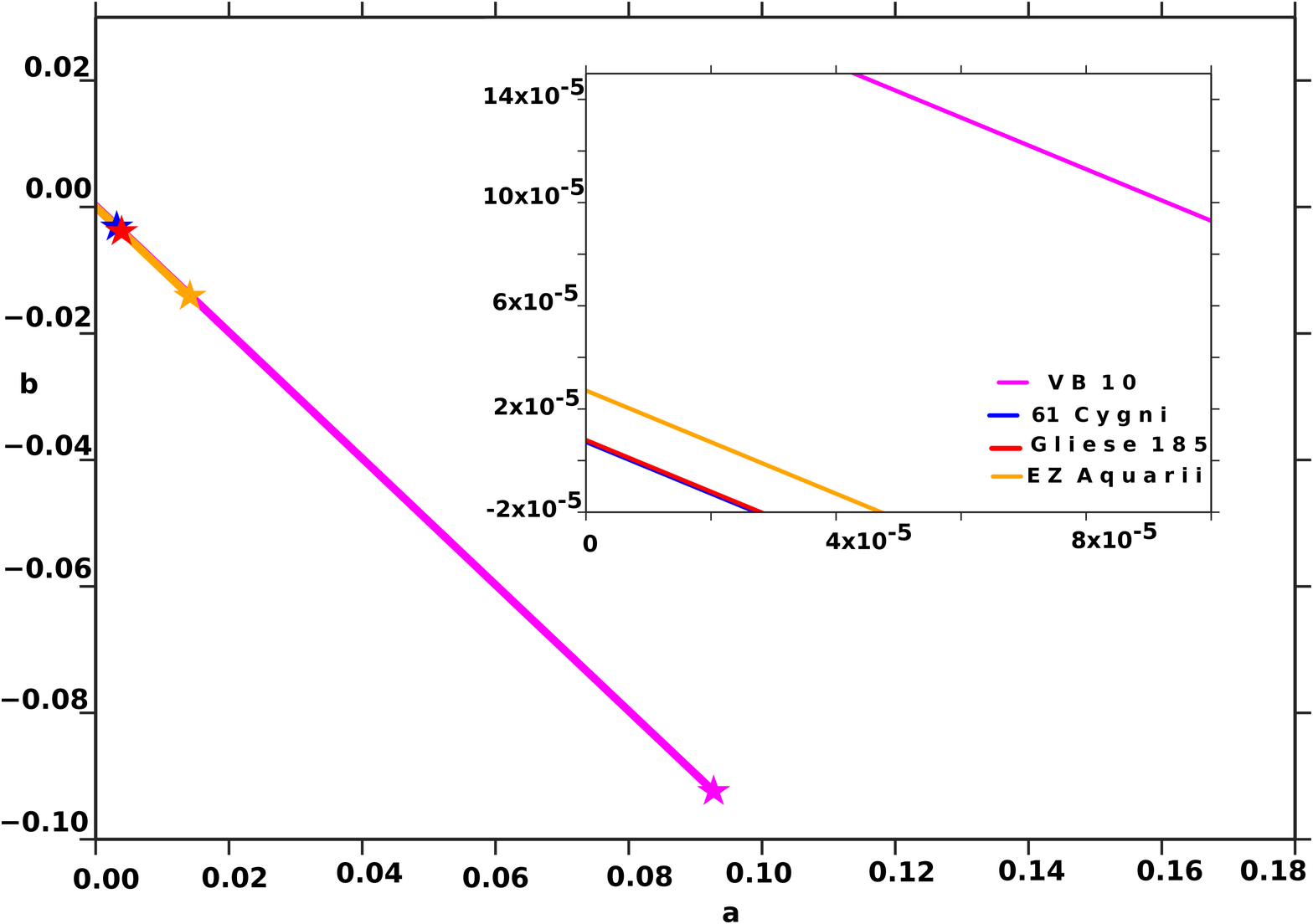}
\caption{Red Dwarfs, the intersection points with the lower bound line of b are shown by '$\star$' s. All of the stars have $n<0.5$.}
\end{figure}
\begin{figure}[h!]
\includegraphics[height=7.5 cm, width=8.417 cm]{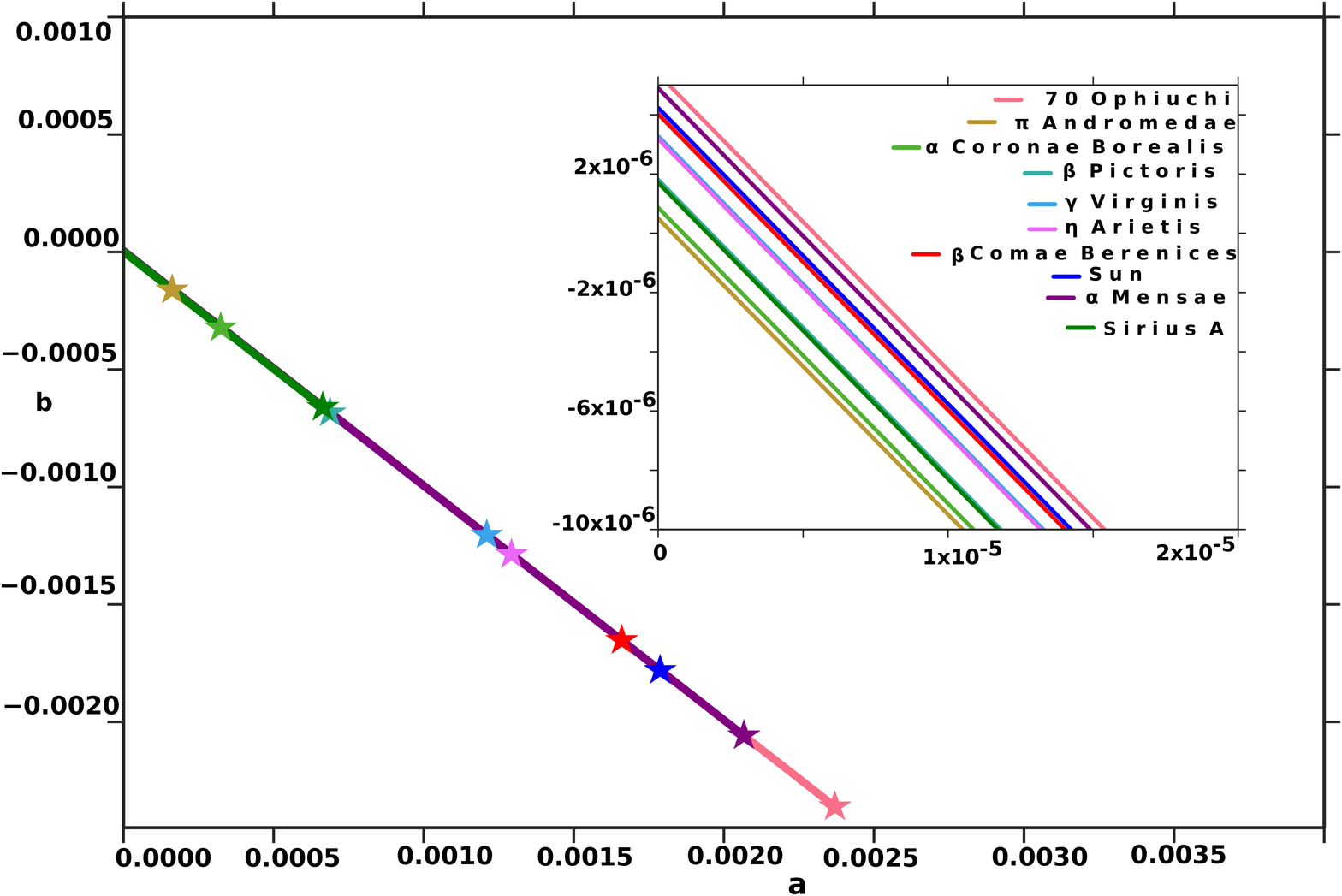}
\caption{Main Sequence stars excluding the red dwarfs, the intersection points with the lower bound line of b are shown by '$\star$' s. All of the stars have $n<0.5$.}
\end{figure}

The stars in the Giant category, are substantially bigger and more luminous than the MS stars or white dwarfs of same temperature. These stars have some sub categories in H-R diagram depending on their temperature or colour and luminosity, i.e; namely, sub giants, bright giants, red giants, yellow giants and blue giants.
\begin{figure}[h!]
\includegraphics[height=7.5 cm, width=8.417 cm]{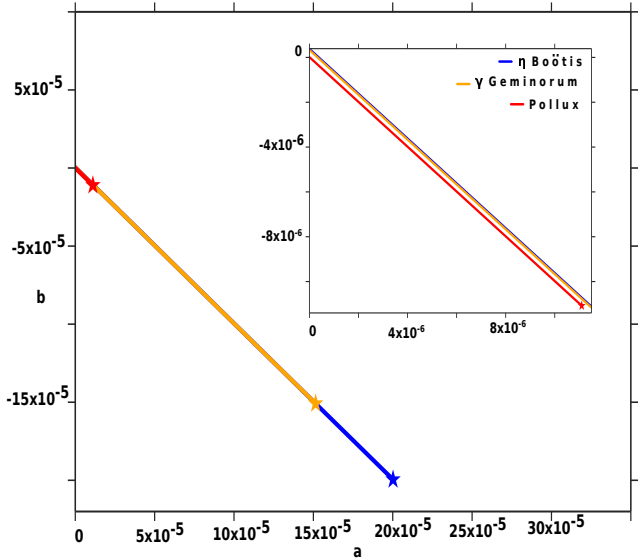}
\caption{Sub giants, the intersection points with the lower bound line of b are shown by '$\star$' s. All of the stars have $n<0.5$.}
\end{figure}
\begin{figure}[h!]
\includegraphics[height=7.5 cm, width=8.417 cm]{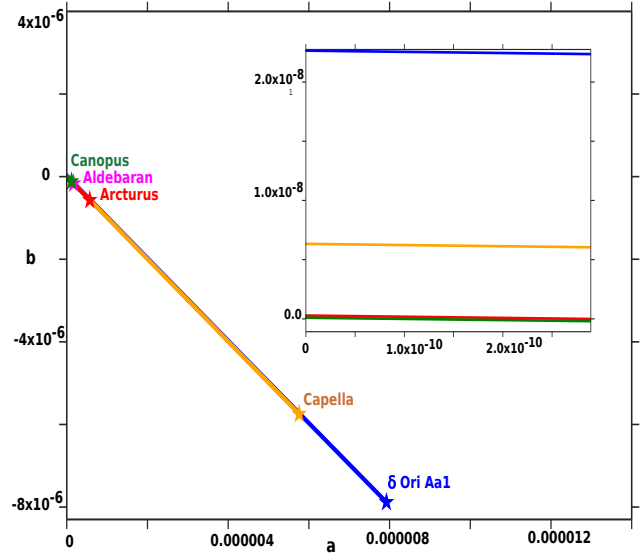}
\caption{Giants other than sub giants and blue giants, the intersection points with the lower bound line of b are shown by '$\star$' s. All of the stars have $n<0.5$.}
\end{figure}

Fig. 24 shows that $a\sim[0,10^{-4}]$ and $b\sim[-10^{-4},10^{-7}]$ for the sub-giants \cite{a19}-\cite{a21} in consideration. Fig. 25 shows that $a\sim[0,10^{-5}]$ and $b\sim[-10^{-5},10^{-8}]$ for the giants \cite{a22}-\cite{a25} other than sub-giants and blue giants.
Fig. 26 shows that $a\sim[0,10^{-4}]$ and $b\sim[-1.5\times10^{-4},0.5\times10^{-6}]$ for the blue-giants\cite{a26}-\cite{a27} in consideration.
\begin{figure}[h!]
\includegraphics[height=7.5 cm, width=8.417 cm]{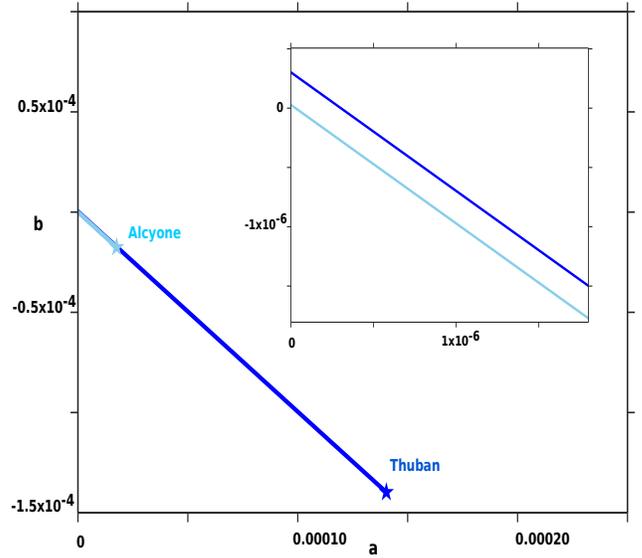}
\caption{Blue or white giants, the intersection points with the lower bound line of b are shown by '$\star$' s. All of the stars have $n<0.5$.}
\end{figure}
\begin{figure}[h!]
\includegraphics[height=7.05 cm, width=8.417 cm]{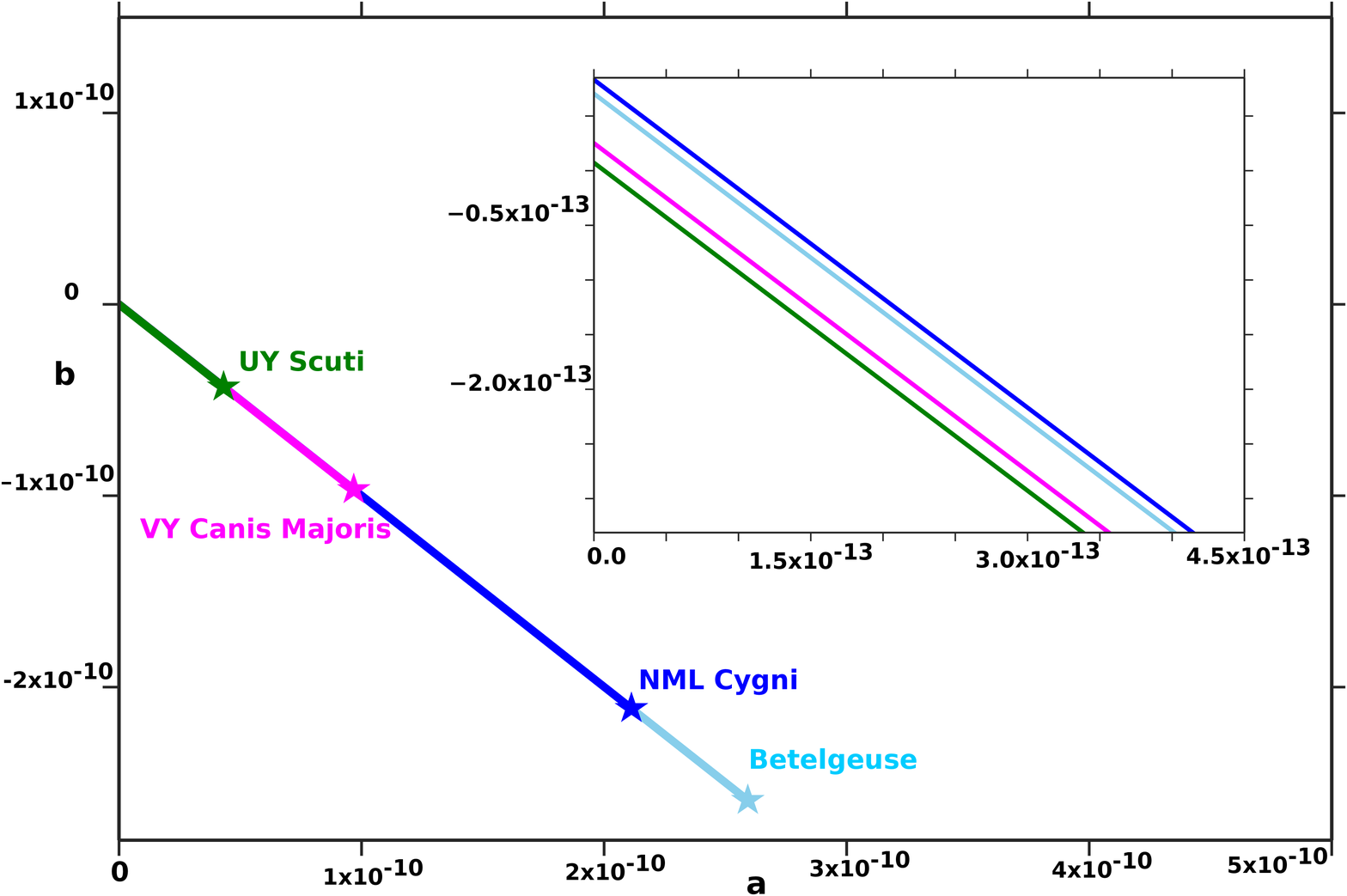}
\caption{Super Giants and Hyper Giants, the intersection points with the lower bound line of b are shown by '$\star$' s. All of the stars have $n<0.5$.}
\end{figure}

Super giants \cite{a29},\cite{a30}are the most luminous and the largest stars in the H-R diagram with luminosity $10^{3}$ to $10^{6}$ $L_\odot$. Super giants have a large range of variation in temperature, i.e; from 4000 K to 40,000 K and also they have a large range in variation in radius, usually from 30 to 500 $R_\odot$, or even in excess of 1,000
R$_\odot$. Hyper giants \cite{a28},\cite{a31},\cite{a32}are unusually big having tremendous luminosity. As for example UY Scuti, currently the largest known star, is in the hyper giant category. Fig. 27 shows that $a\sim[0,10^{-10}]$ and  b   varies within $-10^{-10}$ to $10^{-13}$ for the super-giants and hypergiants in consideration.\\
The above figures reflect the fact that for the white dwarfs, the variation in the parameters, a  and  b  in the feasible region of the a-b parameter space is maximum and for the super giants, that variation is minimum. From the previous section it is clear that for a fixed radius, the more a star is compact, the greater maximum value of a is. Now equation (64) can be written as
\begin{equation}
6(2n-1)x^3+4(5n-1)x^2+13nx+3n=0
\end{equation}
where $x=ar^2$. Hence for a given compactness of a star the upper limit of the parameter, a in the permissible region of a-b parameter space has inverse square relation with the radius of the star.
 From equation (61)
\begin{equation}
b=\frac{n}{r^2}-\frac{(1-n)}{(\frac{1}{a}+r^2)}
\end{equation}
Hence given compactness and radius of a star, the lower limit of b is a monotonically decreasing function of upper limit of a for $n<0.5$. The lower limit of  b  decreases with the maximum permissible value of a and also from Fig. 18 or equation (63) it is obvious that this minimum value of  b  increases with stellar radius. The maximum value of  b  for a star corresponds to $a=0$, this value is $\frac{n}{r^2}$ from equation (62). This value of  b  increases with $n$ and decreases with $r$. As the upper limit of a is a monotonically increasing function of $\frac{n}{r^2}$, hence it is a monotonically increasing function of the upper limit of b. Thus for compact stars having smaller radii have maximum variation in parameter a and b in the permissible region of a-b parameter space. The white dwarfs are the most compact stars having relatively smaller radii in H-R diagram, hence for the white dwarfs the variation in a and  b  are maximum and on the other hand the super giants and hyper giants are the least compact stars having relatively bigger radii, hence for them the variation in a and  b  is minimum. Hence the maximum permissible value of parameter, a or the minimum permissible value of b determines the type of a star.  Table 1 describes mass, radius and compactness of stars of different classes.
\section{Final Remarks and Conclusions}
 The mass and radius of the considered stars in sub-giants class and blue-giant class are of the same order and that is reflected in a-b parameter space too. Obviously, among the considered stars, this model does not distinguish sub-giants and blue giants. Nevertheless, this difference is prominent and clear for the other classes of the considered stars. The considered star has $n=3.18\times 10^{-6}$, as a result the maximum range of parameter, a is 0.0870207 $R_\odot^{-2}$. The compactness of the star and permissible region in a-b parameter space for the star closely matches with the red dwarf category.  Let's consider a brown dwarf, namely, Gliese 229B \cite{a33}, \cite{a34}, it has mass of 0.002 solar mass and radius of 0.047 solar radius, resulting n to be $1.8\times 10^{-7}$. Maximum range of a is 0.166597 $R_\odot^{-2}$ and corresponding minimum value of b is -0.16645 $R_\odot^{-2}$. The radius of the star and the permissible region in a-b parameter space for the star does not match with any of the classes discussed above. Our model clearly distinguishes this brown dwarf from the other stars.  Actually, this model is good in categorising the stars depending on their mass and radius. As discussed in section-10 that, this model clearly classifies stars into two categories, i.e; $0<n<0.5$ and $0.5 \leq n<1$. Given compactness of a star, one of the two above categories for the star is determined. Now one needs an additional input about the star (radius or mass of the star) to  determine on which subcategories, i.e; on which type (discussed in section 11) the star fits in if $n<0.5$ (because all of the stars in H-R diagram have $n<0.5$).  \\
\section{Acknowledgments}
SI is thankful to P Tarafdar of S. N. Bose National Centre For Basic Sciences, for providing some useful insights in the paper. SD is thankful to his sister N Datta for useful discussions about the cubic polynomial appearing in the paper and he is also thankful to his colleague Md. A Shaikh for helping him with the plots.


\newpage
\pagestyle{empty}
 \begin{table*}[t]
\begin{center}
\caption{Stars of several types from H-R diagram}
\begin{tabular}{|c|c|c|c|c|c|c|}
\hline {\bf\color{blue}Stars' type} & \color{blue}\bf Stars' name
  &{\bf\color{blue}M($ \bf M_\odot$)}  &{\bf\color{blue}r ($\bf R_\odot$)  }  & \bf\color{blue} r (km)&  {\bf\color{blue}M (meter)}&$\bf\color{blue}\bf\frac{2M}{r}$ \\ [2ex]

\hline {\bf \color{cyan}White Dwarfs}  & Sirius B  & 0.978	& 0.0084 & 5843.88 & 1442.14902 & 0.0004935587
 \\ [1.5ex]
 & Procyon B & 0.6 & 0.012 & 8348.4 & 884.754 & 0.0002119577 \\ [1.5ex]
 & Van Maanen 2 & 0.68 & 0.011 & 7652.7 & 1002.7212 & 0.0002620568  \\ [1.5ex]
 & 40 Eridani B	& 0.5 & 0.014 & 9739.8 & 737.295 & 0.0001513984 \\ [1.5ex]
& L 97-12 & 0.59 & 0.0128 & 8904.96	& 870.0081 & 0.0001953985 \\ [3ex]
{\bf \color{cyan}Red Dwarfs} & 61 Cygni & 0.69 & 0.74 & 514818 & 1017.4671 & 3.9527$\times 10^{-6}$
\\ [1.5ex]
& Gliese 185 & 0.47 & 0.63 & 438291 & 693.0573 & 3.16254$\times10^{-6}$ \\ [1.5ex]
& EZ Aquarii & 0.21 & 0.32 & 222624	& 309.6639 & 2.781945$\times10^{-6}$ \\ [1.5ex]
& VB 10 & 0.1 & 0.13 & 90441 & 147.459 & 3.26088$\times10^{-6}$ \\ [3ex]
{\bf \color{cyan}MS stars} & $\pi$ Andromedae	& 6.5 & 3.8 & 2643660 & 9584.835 & 7.25118 $\times10^{-6}$ \\ [1.5ex]
& $\alpha$ Coronae Borealis	& 3.2 & 2.5 & 1739250 & 4718.688 & 5.426118$\times10^{-6}$ \\ [1.5ex]
& $\beta$ Pictoris & 2.1 & 1.7	& 1182690 & 3096.639 & 5.23660$\times10^{-6}$ \\ [1.5ex]
& $\gamma$ Virginis	& 1.7 & 1.3 & 904410 & 2506.803 & 5.54351$\times10^{-6}$ \\ [1.5ex]
& $\eta$ Arietis & 1.3 & 1.2 & 834840 & 1916.967 & 4.59241$\times10^{-6}$ \\ [1.5ex]
& $\beta$ Comae Berenices & 1.1 & 1.05 & 730485 & 1622.049 & 0.000004441\\ [1.5ex]
& Sun & 1 & 1 & 695700 & 1474.59 & 4.239154808$\times10^{-6}$ \\ [1.5ex]
& $\alpha$ Mensae & 0.93 & 0.93	& 647001 & 1371.3687 & 4.239154$\times10^{-6}$ \\[1.5ex]
& 70 Ophiuchi & 0.78 & 0.85 & 591345 & 1150.1802 & 0.00000389 \\[1.5ex]
& Sirius A & 2.02 & 1.711 & 1190342.7 & 2978.6718 & 5.004729$\times10^{-6}$\\[3ex]
{\bf \color{cyan}Sub Giants} & $\gamma$ Geminorum& 2.81 & 3.3 & 2295810 & 4143.5979 & 3.6097$\times10^{-6}$ \\[1.5ex]
& $\eta$ Bo\"{o}tis & 1.71 & 2.672 & 1858910.4 & 2521.5489 & 2.7129$\times10^{-6}$\\[1.5ex]
& Pollux & 2.04 & 8.8 & 6122160 & 3008.1636 & 9.8271316$\times10^{-7}$\\[3ex]
{\bf \color{cyan}Giants} & $\delta$ Ori Aa1 & 24 & 16.5 & 11479050 & 35390.16 & 0.000006166
 \\[1.5ex]
& Canopus & 8 & 71 & 49394700 & 11796.72 & 4.7765$\times10^{-7}$\\[1.5ex]
& Arcturus & 1.08 & 25.4 & 17670780 & 1592.5572 & 1.8024$\times10^{-7}$\\[1.5ex]
& Aldebaran & 1.5 & 44.2 & 30749940 & 2211.885 & 1.438627$\times10^{-7}$\\[1.5ex]
& Capella & 2.5687 & 11.98 & 8334486 & 3787.779333 & 9.0894$\times10^{-7}$\\[3ex]
{\bf \color{cyan}Blue Giants} & Alcyone & 3.6 & 8.2	& 5704740 & 5308.524 & 1.86109$\times10^{-6}$\\[1.5ex]
& Thuban & 2.8 & 3.4 & 2365380 & 4128.852 & 3.49106$\times10^{-6}$\\[3ex]
{\bf \color{cyan}Super Giants} & UY Scuti & 8.5 & 1708 & 1.2$\times 10^9$ & 12534.015 & 2.1097$\times10^{-8}$\\[1.5ex]
{\bf \color{cyan}and }& Betelgeuse & 11.6 & 887 & 617085900 & 17105.244 & 5.54388$\times10^{-9}$\\[1.5ex]
{\bf \color{cyan}Hyper Giants}& VY Canis Majoris & 17 & 1420 & 987894000 & 25068.03 & 5.075$\times10^{-8}$\\[1.5ex]
& NML Cygni & 32.5 & 1183 & 823013100 & 47924.175 & 1.1646$\times10^{-7}$\\[1.5ex]
\hline
\end{tabular}
\end{center}
\end{table*}
\end{document}